\documentclass[prl,twocolumn,showpacs,superscriptaddress]{revtex4}
\usepackage{amssymb,amsmath}
\usepackage{graphicx,array}
\usepackage{dcolumn}
\usepackage{bm}

\begin{document}
\title{Superionicity and Polymorphism in Calcium Fluoride at High Pressure}

\author{Claudio Cazorla}
\email{ccazorla@icmab.es}
\thanks{Corresponding Author}
\affiliation{School of Materials Science and Engineering,
             University of New South Wales, Sydney NSW 2052, Australia} 

\author{Daniel Errandonea}
\affiliation{Departamento de F\'isica Aplicada (ICMUV),
             Universitat de Valencia, 46100 Burjassot, Spain}

\begin{abstract}
We present a combined experimental and computational first-principles study
of the superionic and structural properties of CaF$_{2}$ at high $P-T$ 
conditions. We observe an anomalous superionic behavior in the low-$P$ fluorite 
phase that consists in a decrease of the $normal \to superionic$ critical temperature 
with compression. This unexpected effect can be explained in terms of a 
$P$-induced softening of a zone-boundary $X$ phonon which involves exclusively  
fluorine displacements. Also we find that superionic conductivity is absent 
in the high-$P$ cotunnite phase. Instead, superionicity develops in a new 
low-symmetry high-$T$ phase that we identify as monoclinic (space group $P2_{1}/c$).
We discuss the possibility of observing these intriguing phenomena in related isomorphic 
materials.  
\end{abstract}

\maketitle

Alkali earth metal fluorides (i.e., $A$F$_{2}$ compounds with $A$~=~Ca, 
Sr and Ba) blend a family of extraordinary materials which for more than 
six decades have seduced both fundamental and applied physicists. 
Due to their low refractive index, low dispersion and large broadband 
radiation transmittance, $A$F$_{2}$ have been extensively exploited in   
optical devices~\cite{barth90,gan92}. These are also promising candidates 
for solid-state electrolytes to be used in batteries, as their internal electric 
currents dwell on the motion of ions~\cite{hayes85,gillan90,lindan93,maier00,cucinotta09}. 
CaF$_{2}$ is an archetypal ionic conductor and the most representative of $A$F$_{2}$ 
species. Under ambient conditions CaF$_{2}$ crystallizes in the fluorite 
structure ($\alpha$, space group $Fm\bar{3}m$) wherein Ca atoms are cubic 
coordinated to F atoms. In $\alpha$-CaF$_{2}$ a strong increase of the ionic 
conductivity is observed as $T$ is raised up to 
$\sim 1400$~K~\cite{derrington75,hayes89,evangelakis89}. 
At this point the mobility of the fluorine anions is comparable to that of a molten 
salt, while the melting temperature of the system lies $\sim 300$~K 
higher~\cite{mclaughlan67,mitchell72}. 

The accepted dominant effect behind the large ionic conductivity observed in 
$\alpha$-CaF$_{2}$, dubbed as ``superionicity'', is the formation of 
point Frenkel pair defects (FPD) (i.e., simultaneous creation of 
F$^{-}$ vacancies and interstitials)~\cite{gillan90,lindan93}. Yet, 
superionicity requires of abundant and correlated anion displacements thereby 
the formation of FPD by itself appears to be insufficient to fully elucidate the 
atomic mechanisms underlying it~\cite{boyer80,zhou96,schmalzl03}. 
Moreover, this singular transport phenomenon remains surprisingly unexplored at
moderate and high compressions. Almost $40$ years ago, Mirwald and Kennedy were the
first (and to the best of our knowledge the only so far) to experimentally study
the high $P-T$ phase diagram of CaF$_{2}$~\cite{mirwald77}.
By relying on static compression methods and differential thermal analysis, they
reported a continuous reduction of the critical temperature for the
$\alpha$~$\to$~$\beta$ transition (where $\beta$ stands for the superionic phase)
$T_{s}$ under compression. This interesting and apparently puzzling effect has been
overlooked for decades, calling for a meticulous revision with presently improved
experimental techniques. Upon a pressure of $\sim 10$~GPa CaF$_{2}$ 
transforms into the orthorhombic cotunnite phase ($\gamma$, space group 
$Pnma$), which can be structurally related to the cubic $\alpha$-phase through a 
local melting of the F$^{-}$ sublattice~\cite{eddine06}. The detailed ionic processes 
sustaining superionicity in this high-$P$ $\gamma$-phase, however, are totally 
uncertain~\cite{hull98}. As for theory, only few simulation works have recently 
addressed the description of ionic conductivity at non-ambient conditions~\cite{zeng08,cazorla13}. 
Those few computational studies however rely all on molecular dynamics simulations 
performed with semi-empirical pairwise potentials which have been tuned to reproduce the 
behavior of CaF$_{2}$ at ambient pressure. Whether the conclusions attained with 
those models at high $P-T$ conditions may still regarded as reliable, is an 
issue that needs to be clarified. 

\begin{figure}
\centerline{
\includegraphics[width=1.00\linewidth]{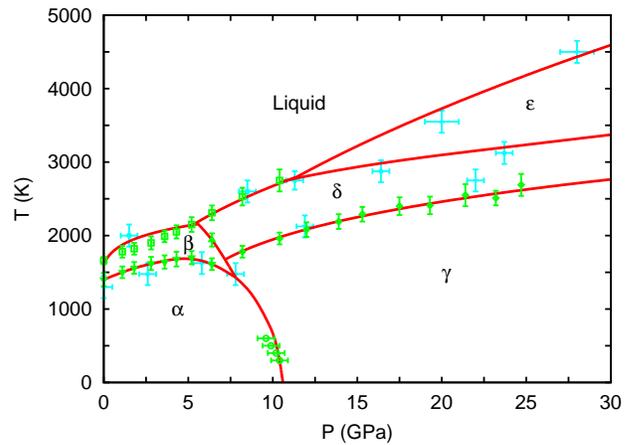}}
\caption{(Color online)~Proposed phase diagram of CaF$_{2}$. Greek
         letters represent normal ($\alpha$, $\gamma$ and $\delta$)
         and superionic ($\beta$ and $\epsilon$) crystal phases, and
         the solid~(red) lines their boundaries.
         Measurements and simulation results are indicated with open~(green)
         and solid~(blue) symbols, respectively.}
\label{fig1}
\end{figure}

In this Letter, we present a combined experimental and first-principles computational 
study of the superionic and structural properties of CaF$_{2}$ at high-$P$ and high-$T$. 
Our observations and simulations disclose an anomalous superionic behavior in 
$\alpha$-CaF$_{2}$ which consists of a decrease of $T_{s}$ in the pressure
interval $5 \le P \le 8$~GPa. We find that such an anomaly can be caused by 
a $P$-induced softening of a zone-boundary $X$ phonon, neither observed in SrF$_{2}$ 
nor in BaF$_{2}$. Regarding $\gamma$-CaF$_{2}$, our first-principles 
simulations show that large ionic conductivity is missing on this phase. 
Rather, superionicity appears after the crystal transforms via a 
second-order transition to a new low-symmetry high-$T$ phase that we 
identify as monoclinic (space group $P2_{1}/c$).  

We carried out two sets of experiments in CaF$_{2}$ [water-free 
99.9~\% purity (Sigma-Aldrich)] using a diamond-anvil cell (DAC). 
In the first set, Raman measurements were performed to determine the 
$\alpha-\gamma$ phase boundary~\cite{speziale02}. In these experiments, 
a $514.5$~nm Ar$^{+}$ laser was used for Raman excitation and silicone-oil 
as a pressure transmitting medium. The temperature was fixed using an electric 
heating sleeve which surrounds the DAC body, and was measured using a 
K-type thermocouple~\cite{errandonea06}. The R$_{1}$ photoluminescence line of 
ruby and the $^{7}$D$_{0}$-$^{5}$F$_{0}$ fluorescence of SrB$_{4}$O$_{7}$:Sm$^{2+}$ 
were used to determine $P$~\cite{mao86,datchi97}. In the second set of 
experiments, CaF$_{2}$ acted as the pressure medium and it was heated using 
the laser-heating technique. A tungsten (W) foil embedded in CaF$_{2}$ was 
heated with a Nd:YLF laser. The sample does neither absorb the laser radiation 
nor emit incandescent light in detectable amounts throughout all the studied  
$P$ range. Therefore, following previous studies on alkali halides~\cite{boehler96}, 
$T$ was measured from the W surface. Structural changes and melting were detected by 
visual observation using the laser speckle technique and a $632.8$~nm He-Ne 
laser~\cite{errandonea00,errandonea13}. Temperatures were determined by fitting 
a Planck function to the thermal emission spectra of W. Series of at least 
four experiments were performed for each transition. The assigned transition
temperature was the average of all the measurements, and the corresponding 
error the maximum absolute deviation. After each heating cycle the DAC was  
inspected to check that neither chemical nor oxidation reactions had 
occurred in the metal surfaces~\cite{errandonea09}. The highest conditions 
reached in our experiments were $T \sim 25$~GPa and $P \sim 2500$~K.  

\begin{figure}
\centerline{
\includegraphics[width=1.00\linewidth]{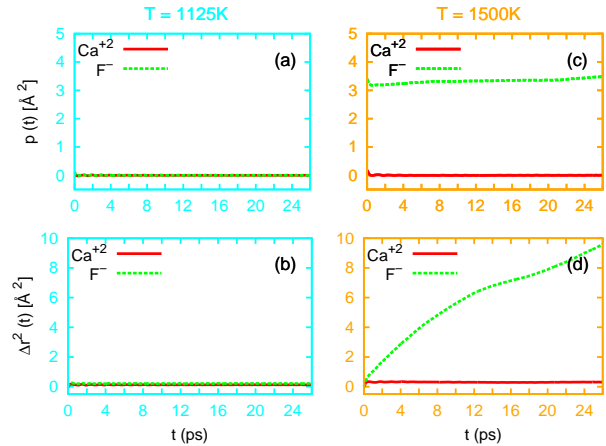}}
\caption{Calculated position correlation function and
         mean squared displacement in CaF$_{2}$ at 
         $T = 1125$~K~(a)-(b) and $1500$~K~(c)-(d) 
         [$P \sim 0$~GPa]. 
         At $1500$~($1125$)~K the system is in the 
	 $\beta$~($\alpha$) phase [see text].}
\label{fig2}
\end{figure}

Our first-principles density functional theory (DFT) calculations were performed
with the generalized gradient approximation to the exchange-correlation energy 
due to Perdew~\cite{pbe96}. We used the ``projector augmented wave'' 
method to represent the ionic cores~\cite{bloch94}, and considered Ca's $6p$-$2s$ 
and F's $2s$-$5p$ electronic states as valence.  
Wave functions were represented in a plane-wave basis truncated at $500$~eV. 
By using these parameters and dense ${\bf k}$-point grids for Brillouin zone 
integrations, we obtained enthalpies converged to within $3$~meV per 
formula unit. In the geometry relaxations, we imposed a force tolerance of 
$0.01$~eV$\cdot$\AA$^{-1}$. The agreement between our zero-temperature DFT 
calculations and the experimental equation of state and Raman frequencies is 
fairly good (see~\cite{supporting}). Details of our one-phase and two-phase 
coexistence \emph{ab initio} molecular dynamics (AIMD) simulations are 
explained in the Supporting Material~\cite{supporting}. 

In Fig.~\ref{fig1}, we show the results of our DAC measurements 
and first-principles simulations in CaF$_{2}$. As it can be appreciated, 
the agreement between experiments and calculations is remarkably good. 
Our results are also in accordance with previous data obtained by 
others at zero pressure [$T_{s} = 1400(90)$~K and the melting temperature 
$T_{m} = 1660(50)$~K]~\cite{derrington75,hayes89,evangelakis89,mclaughlan67,mitchell72}.  
We attempted to estimate the $\alpha-\gamma$ boundary by performing DFT free
energy calculations within the quasi-harmonic approximation however, as we 
will explain later, this approach turns out to be inadequate for CaF$_{2}$.  
The proposed CaF$_{2}$ phase diagram is very rich and complex: five 
different crystal phases [two of which are superionic ($\beta$ and $\epsilon$) 
and two unknown ($\delta$ and $\epsilon$)] and four special three-fold 
coexistence points appear on it. Next, we concentrate on the two findings that 
we deem as the most relevant, namely (i)~observation of anomalous superionic 
behavior in $\alpha$-CaF$_{2}$, and (ii)~appearance of two new high-$T$ phases 
($\delta$ and $\epsilon$) one of which exhibits large ionic conductivity.  

\emph{Anomalous superionic behavior in $\alpha$-CaF$_{2}$}--
Slow motion of the speckle pattern under steady laser illumination 
was ascribed to the $\alpha \to \beta$ transition (where $\beta$ 
stands for the superionic state) in our experiments. 
The measured $\alpha-\beta$ boundary displays a positive variation 
with increasing pressure at low $P$, reaching a maximum 
of $T_{s} = 1700(90)$ at $5.2$~GPa (in stark contrast to data reported
by Mirwald and Kennedy~\cite{mirwald77}). 
At larger compressions, however, d$T_{s}/$d$P$ unexpectedly becomes 
negative (see Fig.~\ref{fig1}). This is in fact a very intriguing 
effect: if superionicity was uniquely mediated by Frenkel pair 
defects (FPD) $T_{s}$ should necessarily increase with raising $P$ because the 
formation energy of FPD escalates with compression (as we show in~\cite{supporting}). 
Indeed, other arguments apart from FPD are needed to satisfactorily 
explain the observed $T_{s}$ anomaly (and probably also superionicity). 

In our AIMD simulations, we identified ionic conductivity by inspecting  
the calculated mean squared displacement $\Delta r^{2}(t)$ of the Ca$^{+2}$ 
and F$^{-}$ ions~\cite{cazorla13}. To ascertain that the crystal remained vibrationally 
stable (i.e., the thermal average position of each ion remains centered on its 
perfect-lattice site), we computed the position correlation function 
$p(t) \equiv \langle [{\bf r_{i}}(t+t_{0}) - {\bf R_{i}}^{0}] 
\cdot [{\bf r_{i}}(t_{0}) - {\bf R_{i}}^{0}] \rangle$ where ${\bf r_{i}}(t)$ 
is the position of ion $i$ at time $t$, ${\bf R_{i}}^{0}$ its perfect-lattice 
position, $t_{0}$ an arbritary time origin, and $\langle \cdot \rangle$ 
denotes thermal average (see Fig.~\ref{fig2}a-c)~\cite{cazorla12,alfe03}. 
The crystal is vibrationally stable if $p(t \to \infty) = 0$ since the  
displacements at widely separated times become uncorrelated. On the contrary, 
if the atoms acquire a permanent vibrational displacement, $p(t \to \infty)$ becomes 
non-zero. As it is shown in Fig.~\ref{fig1}, the results obtained in our $\alpha$-CaF$_{2}$ 
simulations are in very good agreement with our DAC measurements: a negative 
d$T_{s}$/d$P$ slope is found at pressures higher than $\sim 5$~GPa. 
This remarkable agreement between theory and experiments certifies that the 
reported $P$-induced $T_{s}$ anomaly constitutes a genuine effect. 
We note that none of the classical simulation works performed to 
date have reported any such peculiar behavior~\cite{zeng08,cazorla13}, thus we 
may conclude that the employed interaction models are unsuitable to 
emulate CaF$_{2}$ at high $P-T$ conditions. 

\begin{figure}
\centerline{
\includegraphics[width=1.00\linewidth]{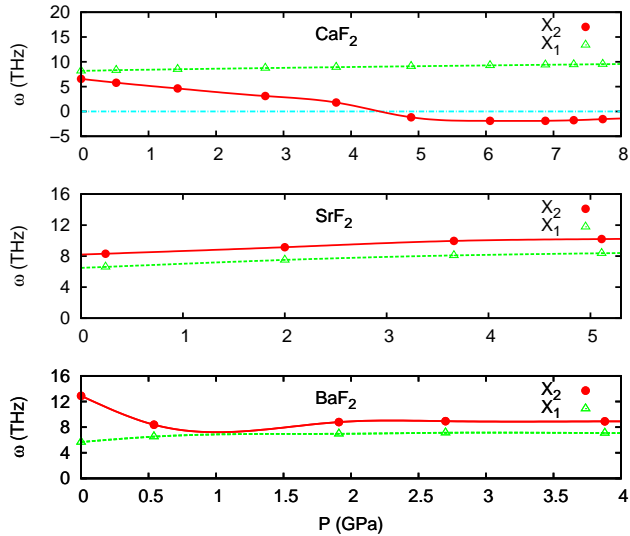}}
\caption{Calculated frequencies of zone-boundary $X$ phonon modes involving 
         F$^{-}$ displacements only, expressed as a function of pressure 
         and $A$F$_{2}$ species (cubic $\alpha$ phase).}
\label{fig3}
\end{figure}

In the search to rationalize the origins of the observed $T_{s}$ anomaly, 
we turned our attention onto collective phonon excitations~\cite{boyer80,zhou96,schmalzl03}.
Recently, an irregular $T$-dependence of a low-energy phonon mode at the $X$ point 
of the Brillouin-zone boundary (single degenerate, labeled here as $X_{2}$) has
been observed in inelastic neutron scattering experiments~\cite{schmalzl03}. 
That irregularity consists in a large decrease of its frequency with increasing 
temperature, which might be related to the onset of superionicity~\cite{boyer80,zhou96,boyer81}. 
Motivated by these findings we studied the $P$-dependence of this and another  
zone-boundary $X$ mode (doubly degenerate, labeled here as $X_{1}$), both of 
which involve only the motion of F$^{-}$ ions, with computational DFT 
methods (see Fig.~\ref{fig3})~\cite{supporting}. Our results show that the $X_{1}$ frequency 
is always the highest and that it increases mildly with compression. The $X_{2}$ mode, by contrast, 
softens significantly with pressure and its frequency eventually becomes imaginary at 
$P \sim 4.5$~GPa. This predicted phonon mode softening marks the appearance of a collective 
instability within the F$^{-}$ sublattice at a pressure that is similar to that identified 
with the $T_{s}$ anomaly. In particular, the $X_{2}$ eigenmode involves rows of anions moving 
anti-phase along an edge of their cubic lattice thus enhancing F$^{-}$ 
disorder~\cite{boyer80,boyer81}. We therefore link the causes behind the observed $T_{s}$ 
anomaly with the predicted $P$-induced $X_{2}$ mode softening in $\alpha$-CaF$_{2}$.  
The results enclosed in Fig.~\ref{fig3} also show that standard quasi-harmonic approaches 
are inadequate to describe CaF$_{2}$ under pressure because of the presence of imaginary 
phonon frequencies which are associated to large anharmonicity. 
Furthermore, aimed at disclosing whether the $T_{s}$ anomaly could be observed also in 
other $A$F$_{2}$ compounds, we carried out analogous phonon calculations in 
$\alpha$-SrF$_{2}$ and $\alpha$-BaF$_{2}$ (see Fig.~\ref{fig3}). Our results show 
that both $X_{1}$ and $X_{2}$ zone-boundary modes behave normally on these materials, 
thus CaF$_{2}$ appears to be unique of its kind. 

\begin{figure}
\centerline{
\includegraphics[width=1.00\linewidth]{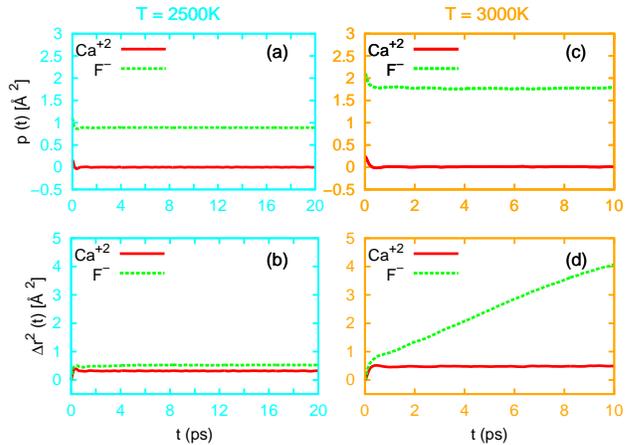}}
\caption{Calculated position correlation function and
         mean squared displacement in CaF$_{2}$
         at $T = 2500$~K~(a)-(b) and $3000$~K~(c)-(d) 
         [$P \sim 17$~GPa].
         At $3000$~($2500$)~K the system is in the 
         $\epsilon$~($\delta$) phase [see text].}
\label{fig4}
\end{figure}

\emph{New high-$T$ $\delta$ and superionic $\epsilon$ phases}--
As $\gamma$-CaF$_{2}$ was stabilized in our DAC experiments and 
$T$ increased steadily, we eventually observed a spatial fluctuation 
of the laser speckle that was markedly different from those found 
in the previous analyzed transitions $\alpha \to \beta$ and 
$\beta \to liquid$. Specifically, an intermittent laser speckle 
motion was detected that could be interpreted as a recrystallization process. 
In order to shed light on the nature of these observations, we performed 
new exhaustive AIMD simulations in $\gamma$-CaF$_{2}$. The analysis of 
our computational results actually shows that the explained detections 
mark the occurrence of a continuous solid-solid transition
(i.e., boundary $\gamma-\delta$ in Fig.~\ref{fig1}). 

In Fig.~\ref{fig4}a-b, we enclose the $\Delta r^{2}$ and $p$ functions
obtained in CaF$_{2}$ at $P \sim 17$~GPa and $T = 2500$~K. 
The perfect-lattice positions entering function $p$ are those of the 
cotunnite $\gamma$ phase determined at zero temperature. 
The results in the figure reveal that at those conditions the 
diffusion of the F$^{-}$ anions is null while the cotunnite $\gamma$ phase becomes 
vibrationally unstable [that is, $\Delta r^{2} (t \to \infty) \sim cte.$ and 
$p(t \to \infty) \neq 0$ for the F$^{-}$ anions]. In particular, the permanent 
displacements acquired by the fluorine anions suggest the happening of a continuous 
transformation from the cotunnite $\gamma$ phase to a new unknown structure that 
we label as $\delta$. We repeated our simulations at different $P-T$ conditions and 
arrived always at the same conclusions. Overall, our DFT estimations are in 
very good agreement with the series of measurements conforming the $\gamma-\delta$ 
boundary (see Fig~\ref{fig1}). By further raising $T$ in our AIMD simulations, we 
found that large ionic conductivity appeared in the new high-$T$ $\delta$ phase (see 
Fig.~\ref{fig4}c-d). This constitutes a very important finding since up to
date it was assumed that superionicity in compressed CaF$_{2}$ appeared in 
the usual cotunnite $\gamma$ phase~\cite{hull98}.
We repeated our simulations at different $P-T$ conditions and determined the 
entire $\delta-\epsilon$ boundary, where $\epsilon$ stands for the new high-$T$ 
superionic phase, up to $\sim 25$~GPa (see Fig.~\ref{fig1}). 
Meanwhile, none similar behavior to these just explained have been reported in 
any previous classical simulation work thus we reaffirm our conclusion that  
currently available CaF$_{2}$ interaction models are not adequate for high $P-T$ analysis. 
             
\begin{figure}
\centerline{
\includegraphics[width=1.00\linewidth]{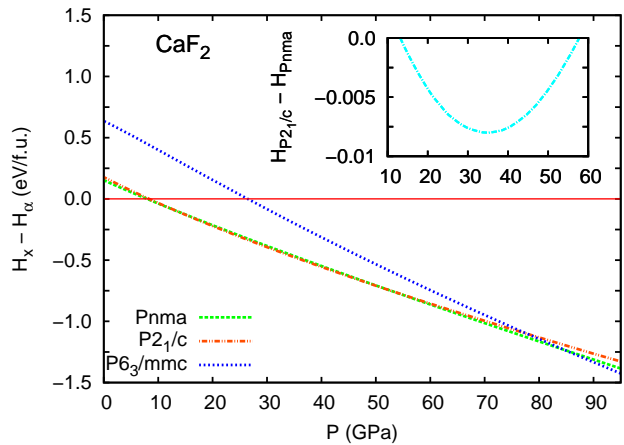}}
\caption{Calculated enthalpy of several crystal structures referred to that
         of the cubic $\alpha$ phase and expressed as a function of pressure.
         \emph{Inset}: Detail of the enthalpy difference between the $P2_{1}/c$ 
         and $Pnma$ phases.}
\label{fig5}
\end{figure}

In our search to identify the symmetry of the newly discovered $\delta$ phase, we initially 
turned our attention onto the hexagonal $P6_{3}/mmc$ phase. We note that
Dorfman \emph{et al.} have recently reported a $T$-induced cotunnite to hexagonal
phase transition in CaF$_{2}$ at pressures higher than $\sim 60$~GPa~\cite{dorfman10}
(see Fig.~\ref{fig5}), hence the $P6_{3}/mmc$ phase seems a natural candidate. 
However, after performing additional AIMD simulations we discarded this because 
the temperatures at which superionicity appeared were below the $\delta-\epsilon$ 
boundary. 
Subsequently, we carried out intensive crystal searches in compressed 
CaF$_{2}$ based on diverse strategies, all which are detailed in the Supporting 
Material~\cite{supporting}. We found several structures which are energetically 
competitive with the $\gamma$ phase at zero temperature, namely: $P2_{1}2_{1}2_{1}$,
$Pmn2_{1}$, $Pc$, $P2_{1}$, and $P2_{1}/c$. 
Of these candidates, we first chose the orthorhombic $P2_{1}2_{1}2_{1}$ and $Pmn2_{1}$, 
and the monoclinic $P2_{1}/c$ phases because a group-subgroup relationship exists between them 
and the orthorhombic $\gamma$ phase (a condition that is required for a continuous  
phase transition~\cite{stokes84,bilbao}). 
Finally, we concentrated in the monoclinic $P2_{1}/c$ phase because this possessed 
the lowest energy. In Fig.~\ref{fig5}, we plot the calculated enthalpy of this monoclinic 
phase in the $0 \le P \le 100$~GPa interval and show that it strongly 
rivals that of the cotunnite structure (actually, at some points both enthalpy curves 
are indistinguishable within the numerical errors). We also computed its phonon 
spectrum at different pressures and found that it is mechanically and vibrationally 
stable (see~\cite{supporting}). Actually, the density of low-energy phonon modes
in the monoclinic $P2_{1}/c$ phase generally is slightly larger than in the orthorhombic 
$Pnma$ phase, which is consistent with a possible $T$-induced phase transition
between the two. As a conclusive test, we performed new exhaustive AIMD 
simulations in the $P2_{1}/c$ phase and found that the calculated superionic 
temperatures were consistent with the $\delta-\epsilon$ boundary previously determined. 
Based on these outcomes, we identify the new high-$T$ $\delta$ phase as monoclinic 
$P2_{1}/c$. This structure has a similar cation coordination polyhedra to that 
of the cotunnite phase: each Ca is linked to $9$ fluorine atoms that form an 
elongated tricapped trigonal prism~\cite{supporting}.        
Furthermore, we performed enthalpy calculations also in SrF$_{2}$ and BaF$_{2}$
under pressure and found that the monoclinic $P2_{1}/c$ phase is likewise 
energetically competitive in these materials (see~\cite{supporting}). Therefore, we 
may conclude that the $\gamma \to \delta$ and $\delta \to \epsilon$ transitions 
observed in CaF$_{2}$ are likely to happen also in SrF$_{2}$ and BaF$_{2}$.   

In summary, based on combined experimental DAC and first-principles simulation 
investigations we have characterized superionic conductivity in archetypal 
CaF$_{2}$ at high $P-T$ conditions, and elucidated further the physical causes 
underlying it. Our findings are relevant also to numerous compounds that are 
isomorphic to CaF$_{2}$ including halides (e.g., PbCl$_{2}$), hydrides 
(e.g., CeH$_{2}$), nitrides (e.g., UN$_{2}$), and simple oxides (e.g., UO$_{2}$). 

This work was supported under the Australian Research Council's
Future Fellowship funding scheme (project number RG134363), and
MICINN-Spain [Grants No. MAT2010-18113, No. CSD2007-00041,
MAT2010-21270-C04-01, CSD2007-00045 and FIS2008-03845]. Computer resources, 
technical expertise and assistance were kindly provided by RES and CESGA. 
We acknowledge helpful comments from two of the referees on the 
proposed CaF$_{2}$ phase diagram.

\end{document}